\documentclass[english,5p]{elsarticle}
\usepackage[T1]{fontenc}
\usepackage[utf8]{inputenc}
\usepackage{color}
\usepackage{float}
\usepackage{amsmath}
\usepackage{graphicx}

\makeatletter

\providecommand{\tabularnewline}{\\}
\newcommand{\lyxdot}{.}

\usepackage{lineno}
\renewcommand\[{\begin{equation}} 
\renewcommand\]{\end{equation}} 

\usepackage[colorlinks,urlcolor=blue]{hyperref}
\usepackage[hyphenbreaks]{breakurl}

\usepackage{tikz-timing}[2014/10/29]
\usetikztiminglibrary[rising arrows]{clockarrows}
\usepackage{xparse} 
\NewDocumentCommand{\busref}{som}{\texttt{%
#3%
\IfValueTF{#2}{[#2]}{}%
\IfBooleanTF{#1}{\#}{}%
}}

\usepackage{tikz}
\usetikzlibrary{shapes.geometric, arrows}
\tikzstyle{process} = [rectangle, rounded corners, text width=2cm, minimum height=2cm, xshift=1cm, text centered, draw=black, fill=white!30]
\tikzstyle{arrow} = [thick, ->, >=stealth]

\usepackage{listings}
\lstdefinelanguage{Cpp}{
  morekeywords={
    JSignal
  },
  morecomment=[l]{//},
}
\lstdefinelanguage{myVhdl}{
  morekeywords=[1]{
    downto, begin, signal, type, signed, array, others
  },
  morekeywords=[2]{
    std_logic_vector
  },
  morecomment=[l]{--},
}

\usepackage{dcolumn}
\newcolumntype{d}[1]{D{.}{.}{#1}}
\newcolumntype{L}[1]{>{\raggedright\let\newline\\\arraybackslash\hspace{0pt}}m{#1}}
\newcolumntype{C}[1]{>{\centering\let\newline\\\arraybackslash\hspace{0pt}}m{#1}}
\newcolumntype{R}[1]{>{\raggedleft\let\newline\\\arraybackslash\hspace{0pt}}m{#1}}

\journal{Nuclear Instruments and Methods in Physics Research A}

\makeatletter
\def\@author#1{\g@addto@macro\elsauthors{\normalsize%
    \def\baselinestretch{1}%
    \upshape\authorsep#1\unskip\textsuperscript{%
      \ifx\@fnmark\@empty\else\unskip\sep\@fnmark\let\sep=,\fi
      \ifx\@corref\@empty\else\unskip\sep\@corref\let\sep=,\fi
      }%
    \def\authorsep{ and }%
    \global\let\@fnmark\@empty
    \global\let\@corref\@empty  
    \global\let\sep\@empty}%
    \@eadauthor={#1}
}
\makeatother

\makeatother

\usepackage{babel}
\usepackage{listings}

\begin{document}

\begin{frontmatter}{}

\title{A software framework for pipelined arithmetic algorithms in field
programmable gate arrays}

\author{J. B. Kim}

\author{E. Won\corref{mycorrespondingauthor}}

\ead{eunil@hep.korea.ac.kr}

\cortext[mycorrespondingauthor]{ Corresponding author}

\address{Physics department, Korea University, Anam-ro 145, Seongbuk-gu, 02841
Seoul, Korea }
\begin{abstract}
Pipelined algorithms implemented in field programmable gate arrays
are extensively used for hardware triggers in the modern experimental
high energy physics field and the complexity of such algorithms increases
rapidly. For development of such hardware triggers, algorithms are
developed in \texttt{C++}, ported to hardware description language
for synthesizing firmware, and then ported back to \texttt{C++} for
simulating the firmware response down to the single bit level. We
present a \texttt{C++} software framework which automatically simulates
and generates hardware description language code for pipelined arithmetic
algorithms.
\end{abstract}
\begin{keyword}
Software framework\sep FPGA\sep Pipelined arithmetic algorithms
\sep VHDL \sep \texttt{C++ }\sep code generation
\end{keyword}

\end{frontmatter}{}


\sloppy

\section{Introduction}

In the modern experimental high energy physics field, detectors with
massive number of channels are used to identify physical processes
that occur when colliding particles. Because the rate of colliding
particles including uninteresting background are in the scale of MHz
\cite{Belle2,CMSCollide,atlasCollide} and data readout from detectors
are in the scale of megabytes \cite{Belle2EventSize,CMSEventSize,atlasEventSize},
it is currently impossible to record all the collision data which
would be produced in the terabyte per second scale. Therefore a hardware
trigger which determines whether the data should be recorded or not
is required. The trigger should filter the detector data in such a
way that only the physics processes of interest are written to a permanent
storage at an acceptable rate. The trigger response also needs to
be prompt in making the decision, because each sub-detector can hold
its data for only a limited amount of time due to hardware limits
which is in the scale of micro seconds \cite{Belle2,CMSLatency,atlasLatency}.
The trigger should perform all of its logic before this limited amount
of time is reached.

Field programmable gate arrays (FPGAs) are integrated circuits that
are programmed using hardware description language. Due to their programmable
and parallel nature, they have been used for event triggers in the
modern experimental high energy physics field \cite{Belle2Trigger,FPGACMS,FPGAATLAS}
extensively.

FPGA based trigger algorithms generally use integer based calculations
\cite{FPGACMS,FPGANN,FPGACMS2,FPGACDF}. Although floating-point calculation
can be implemented in FPGAs, the calculation latency, FPGA resource
usage are significantly higher as discussed in Ref. \cite{FPGAFloat,FPGAFloat2}. 

On the other hand, physics related data are generally handled using
floating-point calculations with general purpose computers and physics
analysis software are built with floating-point calculations for precise
results. One needs to use these software to study the performance
of trigger algorithms. The implemented trigger should also be simulated
in these software in such way that the effects of the trigger on the
recorded physics of interest can be studied, as well as to be compared
with the output from the real hardware down to the single bit level.

Due to these facts, trigger algorithms are usually developed in two
software versions. One that uses floating-point calculations and one
that uses integer calculations \cite{FPGACMS,FPGANN,FPGACMS2,FPGACDF}.
The floating-point version shows the pure algorithm performance while
the integer version shows the degradation of performance due to the
constraints of integer calculation and the performance of the FPGA
algorithms. Due to the coexistence of the two versions, one constantly
needs to synchronize them when the algorithms are modified in one
of the versions. To make matters worse, FPGAs are programmed using
a hardware description language so that there can even be three versions
of the same algorithm that are not necessarily developed by one individual.
These various versions make the maintenance of the level one trigger
software extremely difficult.

A solution to these problems could be to use high level synthesis
(HLS) packages such as Vivado HLS \cite{HLS}. One can write code
in \texttt{C++} and let HLS convert it into a hardware description
language. If latency or resources of a FPGA is not an issue, this
would be the best solution. However, as discussed earlier, floating-point
calculation implemented in FPGAs take up much more latency and resources,
which is also a concern for Vivado HLS \cite{VivadoFloatingPoint}.
Integer or fixed-point algorithms can be written in \texttt{C++} using
the classes provided by Vivado HLS, but then the precision and bit
widths for each calculation should to be additionally considered.
Also there would still be the issue of maintaining two versions, floating-point
and fixed-point, of the same algorithm.

We have developed a framework that solves the multiple version problem
which uses integer based calculations for the hardware description
language. Once an algorithm is implemented in the framework, one can
obtain the floating-point calculation result, integer calculation
result, and very high speed integrated circuit hardware description
language (VHDL) code simultaneously.

In this work, a framework for pipelined arithmetic algorithms in FPGAs
is reported. The goals and design of the framework are explained.
The three \texttt{C++} classes that were developed for the framework
are described. Algorithms that were developed using this framework
are discussed as examples. We also compare between our framework and
Vivado HLS for a linear regression algorithm.

\section{Goals}

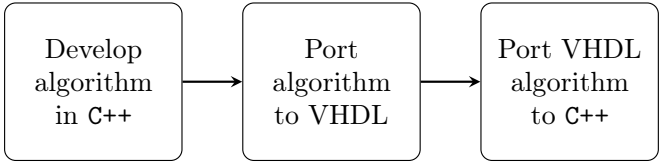
\begin{figure}
\begin{centering}
\resizebox {\columnwidth} {!} {
\begin{tikzpicture}[node distance=2cm]
\node (pro1) [process] {Develop algorithm in \texttt{C++}};
\node (pro2) [process, right of=pro1] {Port algorithm to VHDL};
\node (pro3) [process, right of=pro2] {Port VHDL algorithm to \texttt{C++}};
\draw [arrow] (pro1) -- (pro2);
\draw [arrow] (pro2) -- (pro3);
\end{tikzpicture}
}
\par\end{centering}
\caption{A development procedure for firmware algorithms. An algorithm is developed
with floating-point based calculations using simulated data as input.
It is then ported to VHDL with integer based calculations to synthesize
firmware for a given FPGA. To study the performance of the synthesized
firmware, it is ported back to software which simulates the firmware
response exactly bit-by-bit. \label{fig:FirmwareAlgorithmDevelopment}}
\end{figure}

A typical procedure of firmware algorithm development is shown in
Fig. \ref{fig:FirmwareAlgorithmDevelopment}. Algorithms are developed
and tested in \texttt{C++} first. After the algorithms are validated
they are ported to hardware description language such as VHDL. There
are several issues that should be considered when porting to VHDL.
Floating-point numbers should be converted to integers. The bit width
of all variables should be determined. Division and non-linear operators
such as trigonometric operators should be implemented using look-up
tables (LUTs). The inputs to an operator should be properly buffered
so that the clock cycle between them are in synchronization. Overflow
and underflow should be prevented when doing addition, subtraction,
and multiplication. In order to limit the FPGA resource usage, the
bit width of the inputs to the multiplication operator should be small
enough to be implemented in a digital signal processing (DSP) slice
\cite{DSP}. After porting to VHDL, the resources used by the algorithm
should be small enough to fit in to the chosen target FPGA. One way
to reduce the resource is by controlling the bit widths for the LUTs.
Due to the loss of calculation precision when porting, the VHDL codes
need to be simulated, a priori to confirm if they can achieve their
goals. A floating-point calculation of the algorithm should be performed
to confirm the loss of precision due to this integer conversion. The
VHDL codes should be simulated in \texttt{C++} in such a way that
the results can be used in studying other algorithms. Simulation in
\texttt{C++} will also help in debugging the firmware algorithm most
efficiently.

A framework is developed to simplify the entire process of pipelined
arithmetic firmware algorithm development. The framework can execute
the algorithm using floating-point calculations, simulate the integer-valued
version of the algorithm, automatically generate VHDL code, and deal
with all the issues described previously on arithmetic algorithms.
After an algorithm is developed, the framework will handle the rest
of the development process most efficiently.

\section{Design}

Three classes have been developed in total. The first one is for simulating
VHDL signals and the second one is for LUTs that use block random
access memories (BRAMs) \cite{BRAM}. The third class is to store
the information related with our VHDL codes. Clock cycles are taken
into consideration so that the signals are properly buffered for the
pipelined algorithms in the VHDL code.

\subsection{Signal class}

The signal class has been implemented to simulate the signed and unsigned
VHDL types. Since algorithms generally use floating-point variables
but VHDL signals are integer variables, a conversion from floating-point
values to integer values are executed when the range of the floating-point
variables and bit widths are given. For signed variables, the conversion
is done by the following equations
\begin{align}
\text{symmetric max} & =\text{max}\left(\text{maximum float value },\right.\nonumber \\
 & \left.\left|\text{minimum float value}\right|\right)\\
\text{convertion constant} & =\frac{2^{\left(n-1\right)}-0.5}{\text{symmetric max}}\\
\text{integer varaible} & =\left\lfloor \text{float variable}\right.\nonumber \\
 & \left.\times\text{ convertion constant}\right\rceil ,
\end{align}
where $n$ is a given the bit width, max is the maximum function,
$\left\lfloor \right\rceil $ is a round-off function, and float refers
to a floating-point. For unsigned variables, the conversion is done
by the following equation
\begin{align}
\text{convertion constant} & =\frac{2^{n}-0.5}{\text{maximum float value}}\\
\text{integer variable} & =\left\lfloor \text{float variable}\right.\nonumber \\
 & \left.\times\text{ convertion constant}\right\rceil ,
\end{align}
where $n$ is a given bit width. The real value which the integer
value represents can be calculated using following equation
\begin{equation}
\text{real value}=\frac{\text{integer value}}{\text{convertion constant}}.
\end{equation}
Addition, subtraction and multiplication operators have been implemented
as class methods. The maximum and minimum values are calculated and
stored in the class so that bit widths can be reduced to a minimum
for each operator. Before adding and subtracting, the input's conversion
constants should be matched. They are similarly matched by multiplying
a factor of two which is done by bit shifting. The multiplication
method is implemented so that only one DSP slice is used to reduce
FPGA resources. One DSP slice can perform 25 bit $\times$ 18 bit
calculations so that the bit width of the input is constrained to
25 bits or 18 bits by applying bit shifts. An if-else method is also
implemented to be able to control the flow of the algorithm. It consists
of a comparing component and an assigning component. Two signals can
be compared with a compare method which receives \texttt{==}, \texttt{!=},
\texttt{>=}, \texttt{>}, \texttt{<=}, \texttt{<}, \texttt{\&\&}, and
\texttt{||} as an argument and returns a Boolean type signal. Depending
on the comparison, different arithmetic operations can be preformed
by setting the assigning component. 

Each method for this class has logic which can generate VHDL code.
To reduce calculation overhead, a flag is used to turn it on and off.
All the methods also perform floating-point calculations where the
results are stored in the class so that it can be compared with the
integer-valued calculations.

The \texttt{<=} operator is overloaded to represent that the logic
should be performed in one clock cycle as in VHDL. When this operator
is used, the clock cycle of the signal in the left-hand side will
be assigned with one addition clock cycle compared to the right-hand
side.

An example \texttt{C++} code for pipelined addition using the framework
is shown in Listing \ref{lis:exampleCpp}. After two $\phi$ values
are added together in one clock cycle, another $\phi$ value is added
to the sum in the next clock cycle, as shown in the bottom part of
Listing \ref{lis:exampleCpp}. The automatically generated VHDL code
is shown in Listing \ref{lis:exampleVHDL}. The signals are defined
according to the logic in the implemented \texttt{C++} code. The buffers
required for pipelining the logic are also defined. Sequential VHDL
statements are written according to the \texttt{C++} code. The framework
simulated results is shown in Table \ref{tab:exampleResult}. The
simulated floating-point values, integer values and real values for
each signal are shown. It demonstrates that the framework simulation
is working well.

\inputencoding{latin9}\begin{lstlisting}[caption={Example \texttt{C++} code for a pipelined addition. The \texttt{\textcolor{black}{<=}}
operator is overloaded to represent the logic should be performed
in one clock cycle. \texttt{\textcolor{black}{phi\_0}}, \texttt{\textcolor{black}{phi\_1}},
and \texttt{\textcolor{black}{phi\_2}} are defined as signed signals
with 10 bits and have a range from $-$3.14 to 3.14. \texttt{\textcolor{black}{phi\_0}},
\texttt{\textcolor{black}{phi\_1}}, and \texttt{\textcolor{black}{phi\_2}}'s
current values are 1.57, $-$0.785, and 0.785. \texttt{\textcolor{black}{phi\_0}}
and \texttt{\textcolor{black}{phi\_1}} are added during one clock
cycle to obtain \texttt{\textcolor{black}{phiAdd}}. This is added
with \texttt{\textcolor{black}{phi\_2}} to obtain \texttt{\textcolor{black}{phiAdd2}}
on the next clock cycle.},label={lis:exampleCpp},language={Cpp}]
// Define signals
JSignal phi_0   <= JSignal(10,  1.57 , -3.14, 3.14, 0, storage);
JSignal phi_1   <= JSignal(10, -0.785, -3.14, 3.14, 0, storage);
JSignal phi_2   <= JSignal(10,  0.785, -3.14, 3.14, 0, storage);
// Addition
JSignal phiAdd  <= phi_0  + phi_1;
// Pipelined addition
JSignal phiAdd2 <= phiAdd + phi_2;
\end{lstlisting}
\inputencoding{utf8}
\inputencoding{latin9}\begin{lstlisting}[caption={Automatically generated VHDL for a pipelined addition example. The
framework defines the signals according to the \texttt{\textcolor{black}{JSignal}}
properties in the \texttt{C++} code. There is also a buffer for \texttt{\textcolor{black}{phi\_2}}
to syncronize the clock cycle between \texttt{\textcolor{black}{phiAdd}}
and \texttt{\textcolor{black}{phi\_2}} when calculating \texttt{\textcolor{black}{phiAdd2}}.
The framework writes VHDL according to the logic defined in \texttt{C++}.
},label={lis:exampleVHDL},language={myVhdl}]
-- Define signals
signal phi_0   : signed( 9 downto 0) := (others=>'0');
signal phi_1   : signed( 9 downto 0) := (others=>'0');
signal phiAdd  : signed(10 downto 0) := (others=>'0');
signal phiAdd2 : signed(11 downto 0) := (others=>'0');
type S10D1Array is array(0 downto 0) of signed(9 downto 0); 
signal phi_2_b : S10D1Array := (others=>(others=>'0'));

-- Sequential logic
phiAdd     <= resize(phi_0 ,11)+phi_1;
phiAdd2    <= resize(phiAdd,12)+phi_2_b(0);
phi_2_b(0) <= phi_2;
\end{lstlisting}
\inputencoding{utf8}
\begin{table}[H]
\begin{centering}
\begin{tabular}{|c|R{0.6cm}@{\extracolsep{0pt}.}l|d{3.0}|r@{\extracolsep{0pt}.}l|} \hline  
Name & \multicolumn{2}{c|}{Float value } & \multicolumn{1}{c|}{Integer value}  & \multicolumn{2}{c|}{ Real value}
\tabularnewline \hline  \hline  
\texttt{\textcolor{black}{phi\_0}} & 1&570 & 256 & 1&57153
\tabularnewline \hline  
\texttt{\textcolor{black}{phi\_1}} & $-$0&785 & $-$128 & $-$0&78577
\tabularnewline \hline  
\texttt{\textcolor{black}{phi\_2}} & 0&785 & 128 & 0&78577
\tabularnewline \hline  
\texttt{\textcolor{black}{phiAdd}} & 0&785 & 128 & 0&78577\tabularnewline \hline  \texttt{\textcolor{black}{phiAdd2}} & 1&570 & 256 & 1&57153
\tabularnewline \hline  
\end{tabular}
\par\end{centering}
\caption{Results of the pipelined addition example, where float value is the
floating-point value, integer value is the converted integer value
from the floating-point value, and real value is the value that the
integer value represents. Difference between the floating-point values
and integer representation values are due to the conversion from floating-point
values to integer values. \label{tab:exampleResult}}
\end{table}

\subsection{LUT class}

The LUT class generates LUTs with signal instances as input and output
which can be used for operations that are not directly possible in
VHDL. Division and trigonometric operators can be implemented using
this class. The LUTs are implemented using BRAMs. After the LUT class
is properly set, it can generate a text file which has all the values
to be stored in the BRAM. This text file\footnote{This text file is called COE (coefficient file) within the Xilinx
tools.} is then used with a commercial synthesis tool \cite{BRAMGenerator}.
The input is transformed so that its minimum value is zero which reduces
the BRAM size in certain cases. The output of the BRAM also shares
this property. A constant value is added to get the proper output.
Although this process uses a few clock cycles, it can drastically
reduce the BRAM size in certain cases. This class generates VHDL code
which should be used with a Block Memory Generator IPCORE \cite{BRAMGenerator}
and the generated text file. 

\subsection{VHDL code storage class}

This class stores the entire VHDL generated by the signal class and
LUT class, so that the pipelined arithmetic algorithm can be written
to a VHDL file. Also VHDL syntax for design entities, signal declaration,
and buffers can be optionally added when generating the VHDL file.

\section{Implementation examples}

\begin{figure*}
\begin{centering}
\includegraphics[width=1\textwidth]{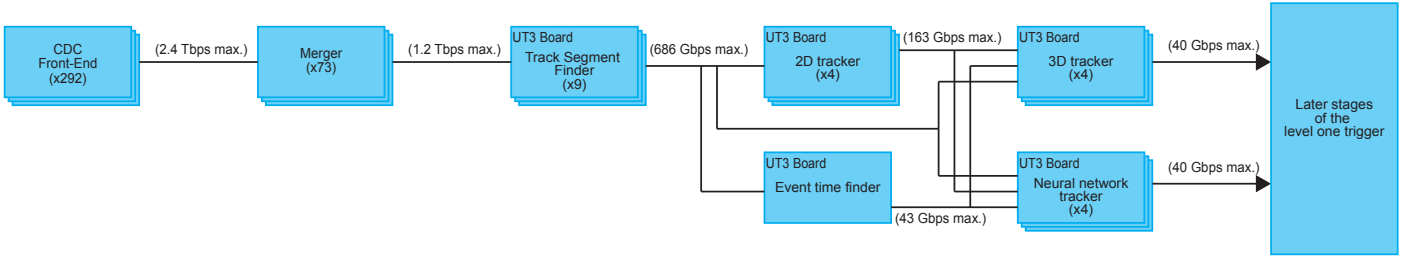}
\par\end{centering}
\caption{Structure of the Belle II level one CDC trigger. The level one CDC
trigger consists (from the left) of CDC front-end boards, merger boards,
track segment finder boards, 2D tracker boards, event time finder
board, 3D tracker boards and neural network boards. Data is transferred
using gigabit optical transceivers between boards. The CDC front-end
boards receive the CDC detector response. The merger boards combine
the CDC front-end data. The track segment finder finds partial tracks.
The 2D tracker find tracks in a two dimensional plane. The event time
finder finds the initial timing of the event. The 3D tracker and neural
network tracker find three dimensions tracks. \label{fig:CDCTRG}}
\end{figure*}

\begin{figure}
\begin{centering}
\includegraphics[width=1\columnwidth]{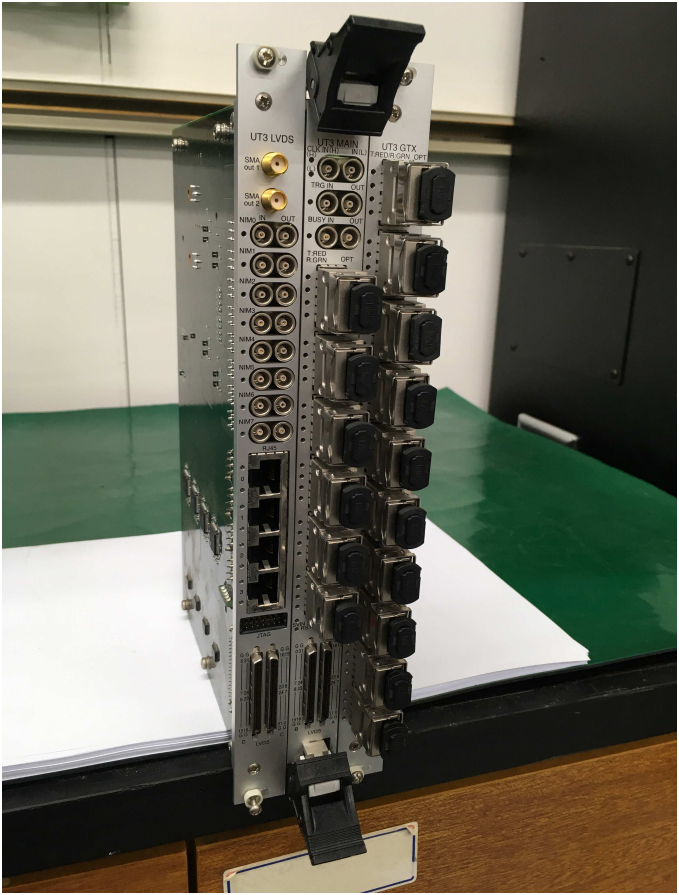}
\par\end{centering}
\caption{A picture of the UT3 board. The UT3 board is a 6U VME board with two
daughter boards that extend the number of channels of communication.
It has a Vertex 6 HXT FPGA  with 40 GTX and 24 GTH gigabit optical
transceivers. It was developed by the high energy accelerator research
organization (KEK) for the Belle II level one trigger. \label{fig:UT3}}

\end{figure}

The Belle II experiment \cite{Belle2} aims to study the charge-conjugation
and the parity violation in $B$ or $D$ meson system precisely and
search for new physics at the SuperKEKB accelerator \cite{SuperKEKB}.
Due to the high beam current and small cross section of physics in
interest, a fast and highly efficient trigger is required. 

The level one trigger is implemented using FPGAs to achieve the above
goals. It consists of several systems, and one of the major systems
is the level one central drift chamber (CDC) trigger. The level one
CDC trigger algorithms are implemented on merger boards \cite{MergerBoard}
and third generation universal trigger boards (UT3), which are 6U
VME \cite{vme} boards with optical cables. The UT3 board was developed
by the high energy accelerator research organization (KEK) for the
Belle II level one trigger. The structure and connections between
the CDC front-end, merger, track segment finder, 2D tracker, 3D tracker
and neural network tracker boards can be seen in Fig. \ref{fig:CDCTRG},
where all connections between the boards are with optical cables.
These UT3 boards which are used for most of the algorithms in the
level one CDC trigger, has a Virtex 6 HXT FPGA with 40 GTX and 24
GTH gigabit optical transceivers which can be seen in Fig. \ref{fig:UT3}. 

The level one trigger uses pipelined algorithms to find patterns potentially
originated from physics of interest. The pipelined algorithms finds
track parameters obtained from the CDC hit information and minimize
$\chi^{2}$ for the track parameter fits. They also include logic
for combining CDC track parameters with electromagnetic calorimeter
(ECL) cluster parameters. Our framework described above has been used
to develop the firmware and \texttt{C++} code for the simulation in
order to implement these algorithms automatically. All the algorithms
in the following sections are implemented on the UT3 boards.

\subsection{$\chi^{2}$ minimization fitters }

There are two fitters that have been developed using our framework.
One fitter minimizes $\chi^{2}$ defined as
\begin{equation}
\chi^{2}=\sum_{i}^{5}\frac{\left[2\left(a\cos\phi_{i}+b\sin\phi_{i}\right)-r_{i}\right]^{2}}{\sigma_{i}^{2}},
\end{equation}
 where $a$ and $b$ are fit parameters and $\phi_{i}$, $r_{i}$
and $\sigma_{i}$ are input variables related with charged tracks
in CDC. The second fitter transforms the wire hit information into
a geometric representation and minimizes a $\chi^{2}$ to obtain track
parameters. The transformation equations are
\begin{align}
\phi_{\text{fineSt}} & =\phi_{st}\pm\text{LUT}\left(\text{TDC}-t_{0}\right)\\
\phi_{\text{ax}} & =\pm\cos^{-1}\left(\frac{r\rho}{2}\right)+\phi_{\text{incident}}\mp\frac{\pi}{2}\label{eq:phiAx}\\
z & =\frac{z_{\text{endplate}}-2r\sin(\frac{\phi_{\text{fineSt}}-\phi_{\text{ax}}}{2})}{\tan\theta_{\text{st}}}\\
s & =\sin^{-1}\left(\frac{r\rho}{2}\right),
\end{align}
where $\phi_{\text{fineSt}}$ is the fine phi position of a hit stereo
wire (wires that have a finite phi shift at the end plates) , $\phi_{\text{st}}$
is the phi position of a hit stereo wire, TDC is the wire hit time
relative to the revolution of the beam, $t_{0}$ is the event time
relative to the revolution of the beam, $\text{LUT}$ is a look up
table that has the x-t curve of the CDC, $\phi_{\text{ax}}$ is the
phi position if a stereo wire is an axial wire (wires that are parallel
to the beam) , $r$ is the radius of a stereo wire layer, $\rho$
is the curvature of a track, $\phi_{\text{incident}}$ is the incident
angle of a track, $z$ is the geometric hit position, $z_{\text{endplate}}$
is the distance from the IP to the end plate of the CDC, and $s$
is the arc length of the track in a two dimensional plane for a stereo
wire layer \cite{3DT}. The $\chi^{2}$ is defined as
\begin{equation}
\chi^{2}=\sum_{i}^{4}\frac{\left[\left(\cot\theta\times s_{i}+z_{0}\right)-z_{i}\right]^{2}}{\sigma_{i}^{2}},
\end{equation}
 where $\cot\theta$ and $z_{0}$ are fit parameters, $z_{i}$ and
$s_{i}$ are the $z$ and $s$ for hit stereo wires, and $\sigma_{i}$
is the resolution of $z_{i}$. There are analytical solutions to these
$\chi^{2}$ minimization which have been used to calculate the fit
parameters. They are 
\begin{align}
a & =\frac{\sum\limits _{i}^{5}\frac{\sin^{2}\phi_{i}}{\sigma_{i}^{2}}\sum\limits _{i}^{5}\frac{r_{i}\cos\phi_{i}}{\sigma_{i}^{2}}-\sum\limits _{i}^{5}\frac{\sin\phi_{i}\cos\phi_{i}}{\sigma_{i}^{2}}\sum\limits _{i}^{5}\frac{r_{i}\sin\phi_{i}}{\sigma_{i}^{2}}}{2\left[^{5}\sum\limits _{i}\frac{\cos^{2}\phi_{i}}{\sigma_{i}^{2}}\sum\limits _{i}^{5}\frac{\sin^{2}\phi_{i}}{\sigma_{i}^{2}}-\left(\sum\limits _{i}^{5}\frac{\sin\phi_{i}\cos\phi_{i}}{\sigma_{i}^{2}}\right)^{2}\right]}\\
b & =\frac{\sum\limits _{i}^{5}\frac{\cos^{2}\phi_{i}}{\sigma_{i}^{2}}\sum\limits _{i}^{5}\frac{r_{i}\sin\phi_{i}}{\sigma_{i}^{2}}-\sum\limits _{i}^{5}\frac{\sin\phi_{i}\cos\phi_{i}}{\sigma_{i}^{2}}\sum\limits _{i}^{5}\frac{r_{i}\cos\phi_{i}}{\sigma_{i}^{2}}}{2\left[\sum\limits _{i}^{5}\frac{\cos^{2}\phi_{i}}{\sigma_{i}^{2}}\sum\limits _{i}^{5}\frac{\sin^{2}\phi_{i}}{\sigma_{i}^{2}}-\left(\sum\limits _{i}^{5}\frac{\sin\phi_{i}\cos\phi_{i}}{\sigma_{i}^{2}}\right)^{2}\right]}
\end{align}
 and 
\begin{align}
\cot\theta & =\frac{\sum\limits _{i}^{4}\frac{1}{\sigma_{i}^{2}}\sum\limits _{i}^{4}\frac{s_{i}z_{i}}{\sigma_{i}^{2}}-\sum\limits _{i}^{4}\frac{s_{i}}{\sigma_{i}^{2}}\sum\limits _{i}^{4}\frac{z_{i}}{\sigma_{i}^{2}}}{\sum\limits _{i}^{4}\frac{1}{\sigma_{i}^{2}}\sum\limits _{i}^{4}\frac{s_{i}^{2}}{\sigma_{i}^{2}}-\left(\sum\limits _{i}^{4}\frac{s_{i}}{\sigma_{i}^{2}}\right)^{2}}\label{eq:cot}\\
z_{0} & =\frac{-\sum\limits _{i}^{4}\frac{s_{i}}{\sigma_{i}^{2}}\sum\limits _{i}^{4}\frac{s_{i}z_{i}}{\sigma_{i}^{2}}+\sum\limits _{i}^{4}\frac{s_{i}^{2}}{\sigma_{i}^{2}}\sum\limits _{i}^{4}\frac{z_{i}}{\sigma_{i}^{2}}}{\sum\limits _{i}^{4}\frac{1}{\sigma_{i}^{2}}\sum\limits _{i}^{4}\frac{s_{i}^{2}}{\sigma_{i}^{2}}-\left(\sum\limits _{i}^{4}\frac{s_{i}}{\sigma_{i}^{2}}\right)^{2}}.\label{eq:z0}
\end{align}
These solutions consist of addition, subtraction, multiplication,
division and trigonometric operations. Division and trigonometric
operations are implemented using LUT class. 

For the second fitter, we used the VHDL code from our framework to
generate firmware. Xilinx ISE \cite{ISE} was used to generate the
firmware for the FPGA on the UT3 board where the clock frequency time
constraint was set to 127 MHz. Xilinx ISE reported that 2,365 slice
registers, 2,919 slice LUTs, 18 RAMB36E1s, 5 RAMB18E1s, and 52 DSP48E1s
were used for the design and that all timing constraints were met.
A firmware based test bench was developed to record the results from
the firmware module that has the generated VHDL code. We compare the
firmware results with the simulated results from our framework. 

\begin{figure}
\begin{centering}
\includegraphics[width=1\columnwidth]{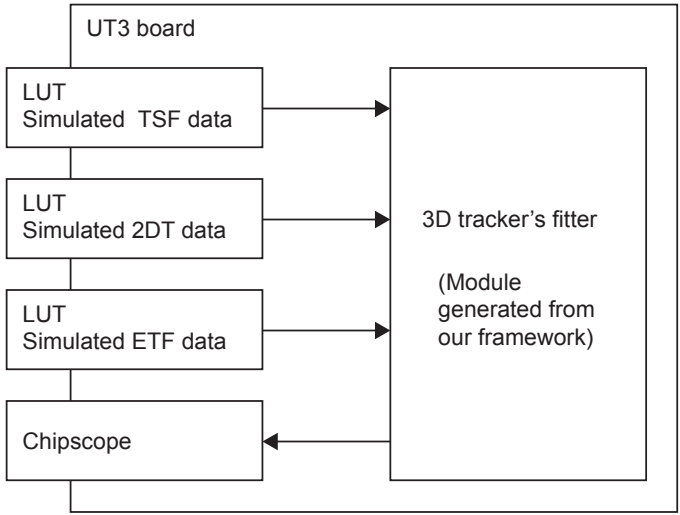}
\par\end{centering}
\caption{\textcolor{black}{Firmware based test bench structure for testing
the 3D tracker's fitter firmware module which was generated from the
framework. The firmware module is tested with data from a TSIM which
are held in LUTs. The LUTs contain tracks segment finder (TSF), 2D
tracker (2DT), and event time finder (ETF) data. Chipscope is connected
to record the output of the firmware.\label{fig:TestBenchStructureDetail}}}
\end{figure}

\begin{figure}
\begin{centering}
\includegraphics[width=1\columnwidth]{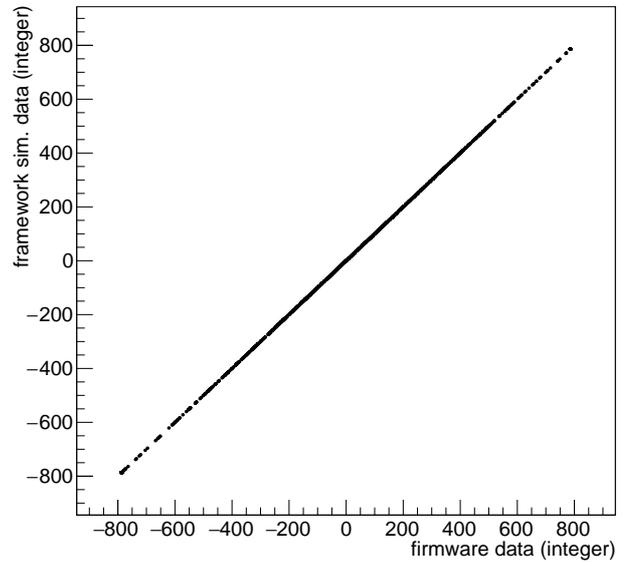}
\par\end{centering}
\caption{Comparison of $z_{0}$ between the recorded firmware results and the
integer based simulation results from the framework (framework sim.)
for an automatically generated VHDL code. The results are exactly
equal. \label{fig:CompareResult}}
\end{figure}

\begin{figure}
\begin{centering}
\includegraphics[width=1\columnwidth]{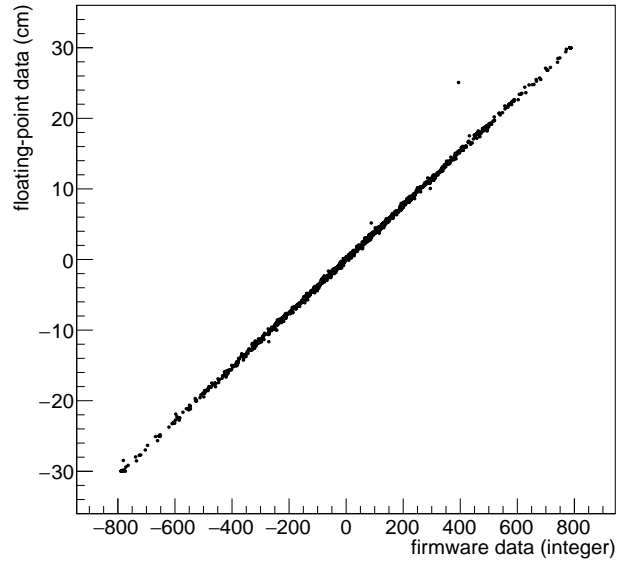}
\par\end{centering}
\caption{Correlation of $z_{0}$ between the recorded firmware results and
the floating-point calculation results. They have a strong correlation
with each other. The outliers are due to the accumulation of lost
precision when processing the pipelined integer based algorithm. \label{fig:CompareFloatResult}}
\end{figure}

\begin{figure}
\begin{centering}
\includegraphics[width=1\columnwidth]{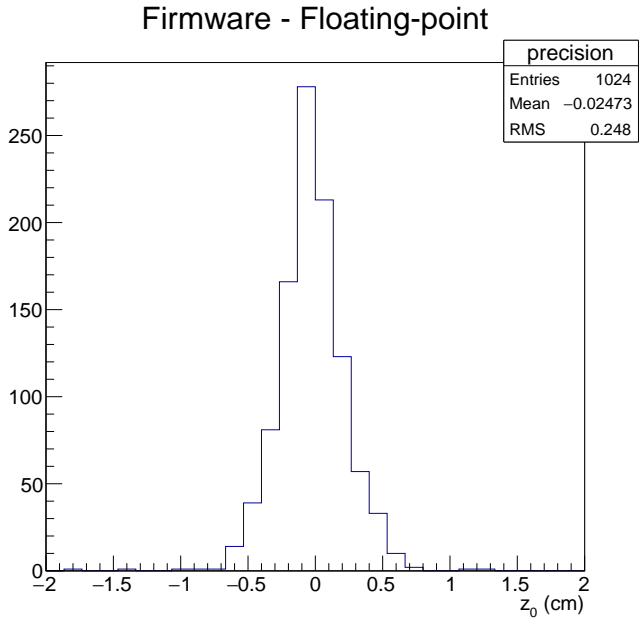}
\par\end{centering}
\caption{Histogram of the firmware precision for $z_{0}$. A conversion constant
was multiplied to the recorded firmware results and subtracted with
the float-point calculated results to calculate the precision. The
RMS is 0.2 cm which is below the expected resolution from the $z_{0}$
fitter algorithm of $\mathcal{O}\left(1\right)$ cm \cite{3DT}. \label{fig:FloatResolution}}
\end{figure}

The structure of the firmware based test bench can be seen in Fig.
\ref{fig:TestBenchStructureDetail}. The firmware based test bench
has LUTs which contain values that are acquired from a trigger simulation
(TSIM). The values are given to the firmware module clock by clock.
The output of the firmware module is connected to Chipscope \cite{Chipscope}
to record the firmware response which is shown in Fig. \ref{fig:ChipscopeResult}.
The values between the recorded the firmware results and simulation
results from the framework are found to be identical down to single
bit level. In Fig. \ref{fig:CompareResult}, a large statistics of
the integer output of the framework and recorded firmware results
for $z_{0}$ are shown to be exactly the same. We were also able to
confirm that the firmware latency is 19 clock cycles which was the
expected value from the framework. The recorded firmware results and
the float-point calculation results from the framework for $z_{0}$
have a strong correlation which is shown in Fig. \ref{fig:CompareFloatResult}.
To see the precision of the recorded firmware results, they are multiplied
with a conversion constant and then subtracted with the float-point
calculation results from the framework where a histogram of $z_{0}$
precision is shown in Fig. \ref{fig:FloatResolution}. The histogram's
root mean square (RMS) was found to be 0.2 cm which is below the expected
resolution from the $z_{0}$ fitter algorithm of $\mathcal{O}\left(1\right)$
cm \cite{3DT}. These results show that our framework works well and
satisfy the level 1 trigger requirements of Belle II.

\begin{figure}
(a)\\
\begin{tikztimingtable}[%
    timing/dslope=0.1,
    timing/.style={x=\columnwidth/11,y=2ex},
    x=\columnwidth/11,
    timing/rowdist=3ex,
    timing/coldist=4pt,
    timing/name/.style={font=\sffamily\scriptsize}
]
\busref{$z_{0}$} & u D{225} D{-435} D{184} D{79} D{630} D{-621} D{-109} D{-528} u\\ 
\busref{$\cot \theta$} & u D{244} D{94} D{23} D{110} D{-22} D{405} D{347} D{350} u  \\ 
\end{tikztimingtable}\\
\bigbreak
(b)\\
\scriptsize
\texttt{memory\_initialization\_radix=10;\\
memory\_initialization\_vector=225,-435,184, 79,630,-621,-109,-528,\\}
\normalsize 
\bigbreak
(c)\\
\scriptsize
\texttt{memory\_initialization\_radix=10;\\
memory\_initialization\_vector=244,\ \ 94, 23,110,-22, 405, 347, 350,\\}

\caption{Recorded results of the firmware using Chipscope and integer based
simulation results from the framework for an automatically generated
VHDL code. In (a), recorded firmware results for $z_{0}$ and $\cot\theta$
are shown. In (b), integer based simulation results from the framework
for $z_{0}$ are shown. In (c), integer based simulation results from
the framework for $\cot\theta$ are shown. The firmware results and
integer based simulation results from the framework match perfectly.
\label{fig:ChipscopeResult}}
\end{figure}

\subsection{CDC geometry calculation}

By using track parameters, the position of the track at a specific
layer of the CDC is required to be calculated. The position is calculated
in two steps. In the first step, our algorithm calculates the $\phi_{\text{ax}}$
position of the track for the layer of the CDC using Eq. \ref{eq:phiAx}.
The second step converts the $\phi_{\text{ax}}$ position to a corresponding
wire position at a given layer of interest. These calculations consist
of addition, subtraction, multiplication and trigonometric operations.
The trigonometric operations were implemented using the LUT class.
A firmware test bench confirmed that the synthesized firmware and
simulated algorithm using the framework return the same results.

\subsection{Combining CDC track and ECL cluster parameters}

The framework has been used in algorithms that combines CDC trigger
and ECL trigger information. The level one CDC trigger outputs track
momentum parameters while the level one ECL trigger outputs cluster
positions created by the deposited energy from the tracks. The two
information can be combined using the position of the track which
can increase the performance of the trigger. Using the track momentum
parameters from the CDC trigger, the expected position of the track
in the ECL detector is calculated which is used to calculate distance
between the expected position and the actual cluster position. This
distance can be used to relate the CDC tracks and ECL clusters. The
ratio between energy and momentum of the track is also calculated
which can help to identify the particle. All of these calculations
are implemented using the developed framework.

\section{Comparison with Vivado HLS}

\begin{table}
\begin{centering}
\begin{tabular}{|c|c|c|c|}
\hline 
Linear regression & float HLS & fixed HLS & framework\tabularnewline
\hline 
\hline 
LUT & 8061 & 1104 & 723\tabularnewline
\hline 
FF & 8064 & 501 & 283\tabularnewline
\hline 
BRAM & 0 & 5 & 2.5\tabularnewline
\hline 
DSP & 126 & 40 & 29\tabularnewline
\hline 
Latency (clocks) & 33 & 5 & 5\tabularnewline
\hline 
$z_{0}$ precision (cm) &  & 0.12 & 0.13\tabularnewline
\hline 
\end{tabular}
\par\end{centering}
\caption{Comparison between HLS and the framework for a linear regression algorithm.
A floating-point calculation (float HLS) and fixed-point calculation
(fixed HLS) version was developed. The resource usage, latency, and
precision are compared, where LUT are slice look-up tables, FF are
flip-flops, BRAM are block RAMs, DSP are DSP slices, latency are the
number of clock cycles to preform the algorithm, and $z_{0}$ precision
is the RMS of a histogram where the results are subtracted with floating-point
calculation results. \label{tab:ComparisonHLS}}

\end{table}

Our framework was compared with Vivado HLS for a linear regression
algorithm (Eq. \ref{eq:cot} and Eq. \ref{eq:z0}). Floating-point
and fixed-point versions were developed using Vivado HLS. For a fair
comparison (latency wise), a LUT that replaces the division operator
was also developed for the Vivado HLS fixed-point version case. All
firmware were synthesized and implemented using the Vivado design
suite \cite{Vivado}. The target FPGA and clock frequency timing constraint
were set to xcvu080-ffvb2104-2-e and 127 MHz. Simulation results from
Vivado HLS and our framework were used to measure the precision of
the firmware where the input data is from TSIM. The precision is measured
by filling a histogram with simulated firmware results subtracted
by floating-point calculated results. We define the precision to be
the RMS of the histogram. The resource usage, latency and precision
can be found in Table. \ref{tab:ComparisonHLS}. We find that within
the expected resolution from the $z_{0}$ fitter algorithm of $\mathcal{O}\left(1\right)$
cm, that our framework uses the least resources and latency, as demonstrated
clearly in Table. \ref{tab:ComparisonHLS}.

\section{Conclusions}

A framework that allows automatic generation of pipelined algorithms
in VHDL is implemented which also simulate the algorithms. It was
validated with $\chi^{2}$ minimization, a sub-detector geometry calculation,
and combining algorithms of sub-detectors. It was compared with Vivado
HLS and our framework is found to be more efficient resource and latency
wise. Development and maintenance of pipelined arithmetic firmware
algorithms using this framework is applied to a variety of situations
and is demonstrated that the framework we developed is most efficient
in dealing with these tasks. Our framework can be used for future
trigger development in an efficient way.

\section*{Acknowledgment}

We acknowledge support from the National Research Foundation of Korean
Grants No. NRF-2017R1A2B3001968.

\bibliographystyle{elsTest}
\bibliography{vhdlCppPaper_v2_6}

\end{document}